           \def\OldLaTeX{0}  

\ifnum\OldLaTeX=1 
    \documentclass[12pt,letterpaper]{article}
\else
    \documentstyle[12pt]{article}
\fi

\newcommand{\r}{\vec{r}}
\newcommand{\R}{\vec{R}}
\newcommand{\runo}{\vec{r}_1}
\newcommand{\rdue}{\vec{r}_2}
\newcommand{\runop}{\vec{r}_1^{~\prime}}
\newcommand{\rduep}{\vec{r}_2^{~\prime}}
\newcommand{\ro}{\vec{r}_0}
\newcommand{\dr}{\dot{\vec{r}}}
\newcommand{\dR}{\dot{\vec{R}}}
\newcommand{\druno}{\dot{\vec{r}_1}}
\newcommand{\drdue}{\dot{\vec{r}_2}}
\newcommand{\drunop}{\dot{\vec{r}_1^{~\prime}}}
\newcommand{\drduep}{\dot{\vec{r}_2^{~\prime}}}
\newcommand{\ddr}{\ddot{\vec{r}}}
\newcommand{\ddR}{\ddot{\vec{R}}}
\newcommand{\ddruno}{\ddot{\vec{r}_1}}
\newcommand{\ddrdue}{\ddot{\vec{r}_2}}
\newcommand{\ri}{\vec{r}_i}
\newcommand{\rf}{\vec{r}_f}
\newcommand{\ra}{\vec{r}_A}
\newcommand{\rb}{\vec{r}_B}
\newcommand{\runoo}{\vec{r}_{10}}
\newcommand{\rdueo}{\vec{r}_{20}}
\newcommand{\xuno}{x_1}
\newcommand{\xdue}{x_2}
\newcommand{\xo}{x_0}
\renewcommand{\xi}{x_i}
\newcommand{\xf}{x_f}
\newcommand{\xa}{x_A}
\newcommand{\xb}{x_B}

\newcommand{\yuno}{y_1}
\newcommand{\ydue}{y_2}
\newcommand{\yo}{y_0}
\newcommand{\yi}{y_i}
\newcommand{\yf}{y_f}
\newcommand{\ya}{y_A}
\newcommand{\yb}{y_B}

\newcommand{\vuno}{{\vec{v}_1}}
\newcommand{\vunop}{{\vec{v}_1^{~\prime}}}
\newcommand{\vdue}{{\vec{v}_2}}
\newcommand{\vduep}{{\vec{v}_2^{~\prime}}}

\renewcommand{\u}{\vec{U}}

\renewcommand{\t}{\tau}
\renewcommand{\to}{t_0}
\newcommand{\tb}{\bar{t}}

\newcommand{\tfi}{(t_f-t_i)}
\newcommand{\toi}{(t_0-t_i)}
\newcommand{\tbto}{(\bar {t}+\tau -t_0)}
\newcommand{\tob}{(t_0-\bar {t})}
\newcommand{\tfo}{(t_f-t_0)}

\newcommand{\roi}{(\vec{r}_0-\vec{r}_i)}
\newcommand{\rob}{(\vec{r}_0-\vec{r}_B)}
\newcommand{\roa}{(\vec{r}_0-\vec{r}_A)}
\newcommand{\rof}{(\vec{r}_0-\vec{r}_f)}
\newcommand{\xoi}{(x_0-x_i)}
\newcommand{\xob}{(x_0-x_B)}
\newcommand{\xoa}{(x_0-x_A)}
\newcommand{\xof}{(x_0-x_f)}
\newcommand{\yoi}{(y_0-y_i)}
\newcommand{\yob}{(y_0-y_B)}
\newcommand{\yoa}{(y_0-y_A)}
\newcommand{\yof}{(y_0-y_f)}

\renewcommand{\v}{\vert}
\renewcommand{\d}{\delta}
\renewcommand{\l}{\lambda}
\newcommand{\E}{{\cal E}}

\newcommand{\p}{\prime}
\newcommand{\non}{\nonumber}
\newcommand{\rff}[1]{(\ref{#1})}
\newcommand{\we}{\wedge}
\newcommand{\pa}{\partial}
\newcommand{\lb}{\label}

\begin{document}

\thispagestyle{empty}

\addtocounter{footnote}{1}
\renewcommand{\thefootnote}{\fnsymbol{footnote}}

\vspace{-20pt} 



\vspace{18pt} 

\begin{center}
{ {\Large {\bf
Time machines: \\ the Principle of Self-Consistency \\
as a consequence of \\ the Principle of Minimal Action}}}

\vspace{18pt}

A. Carlini\\ 
{\small\em {NORDITA, Blegdamsvej 17, DK-2100 Copenhagen \O ,
Denmark}}\\[1.2em] 

V. P. Frolov\\ 
{\small\em {Theoretical Physics Institute, University of 
Alberta, Edmonton, Canada T6G 2J1}} \\
{\small\em {P.N. Lebedev Physical Institute, Leninsky Prospect 53,
Moscow, Russia}}\\[1.2em] 

M. B. Mensky\\ 
{\small\em {P.N. Lebedev Physical Institute, 117924, Moscow,
Russia}}\\
{\small\em {TAC, Blegdamsvej 17, DK-2100 Copenhagen \O ,
Denmark}} \\[1.2em]

I. D. Novikov\\ 
{\small\em {The Copenhagen University Observatory, 
\O stervoldgade 3, DK-1350 Copenhagen K, Denmark}}\\ 
{\small\em{NORDITA, Blegdamsvej 17, DK-2100 Copenhagen \O ,
Denmark}} \\
{\small\em {TAC, Blegdamsvej 17, DK-2100 Copenhagen \O ,
Denmark}} \\
{\small\em {Astro Space Center of the P.N. Lebedev Physical
Institute, Profsoyuznaja 84/32, Moscow,}}\\
{\small\em {117810, Russia}}\\[1.2em] 

H. H. Soleng\\ 
{\small\em {Theory Division, CERN, CH-1211 Geneva 23, Switzerland}}\\ 
\vspace{24pt}
\end{center}

\newpage 

{\centerline{\bf Abstract}}

\bigskip
We consider the action principle to derive the classical, non-relativistic
motion of a self-interacting particle in a 4-D Lorentzian spacetime containing
a
wormhole and which allows the existence of closed time-like curves.
For the case of a `hard-sphere' self-interaction potential we show that
the only possible trajectories (for a particle with fixed initial and final
positions and which traverses the wormhole once) minimizing the classical
action are those which are globally self-consistent, and that the `Principle
of self-consistency' (originally introduced by Novikov) is thus a natural
consequence of the `Principle of minimal action.'

\newpage

\addtocounter{footnote}{-\value{footnote}}
\renewcommand{\thefootnote}{\alph{footnote}}
\section{Introduction}

The possibility that the laws of physics might allow for the
existence of closed time-like curves (CTCs) inside our universe has been a
long time conjecture \cite{god}, which
has been more recently revived by a series of papers [2--15].
Macroscopic CTCs might be easily realized as a semiclassical consequence of
the `quantum foam' structure of spacetime at Planck scales (see, e.g.,
Ref.~\cite{haw}).
The idea that 4-d geometry itself might be no longer a
fundamental concept, and that  close to the Planck scale 
one should instead
allow for a `quantum fuzz' in which spacetime continuosly undergoes
non-trivial topological fluctuations, was first introduced by Wheeler
\cite{whe}.

One peculiar kind of these topological fluctuations is the so-called
wormhole, intuitively speaking a 4-d `handle-like' geometry, whose
two `mouths' join distant regions of spacetime.
Provided the matter density in some regions of our universe
satisfies certain properties (i.e.\ it violates
the so-called `averaged weak energy conditions,' see, e.g., Refs.~[3--6]),
these 4-d Lorentzian wormhole geometries could, in principle, exist
(at least they can be exact solutions of the Einstein equations).

It has then been shown [4, 7--8]
that generic relative motions of the two
wormhole's mouths, or equivalently generic gravitational redshifts at the mouths
due to external gravitational fields, can indeed in principle produce
CTCs: if the wormhole is traversed from mouth to mouth, it acts as
a `time machine' allowing one to travel into the past or
into the future.

For spacetimes with CTCs, past and future are no longer `globally'
distinct, and the (Cauchy) problem of evolving the equations of motion of a
particle (or field) from a set of initial conditions into the `future'
is in general more involved than for spacetimes without CTCs [9, 11, 12].
In particular, as originally pointed out in Ref.~\cite{nov1},
events on CTCs should causally influence each other along the
`loops in time' in a self-adjusted, consistent way.
This requirement has been explicitly formulated
as the `Principle of self-consistency,' according to which
{\it the only solutions to the laws of physics that can occur
locally in the real universe are those which are globally
self-consistent} [7, 9, 13--15].

The question arises whether the `Principle of self-consistency'
really is an independent assumption which is necessary in order
to make sense of the spacetimes with CTCs without resorting to
`new physics,' or whether it 
actually 
can be 
incorporated
into some other, more fundamental, physical principle.

In this paper we consider, in the framework of the action principle,
the problem of the classical, non-relativistic
motion of a self-interacting point particle, passing
once trough a wormhole `time machine.'
For this idealized model we are able to show that the `Principle
of self-consistency' is in fact a direct consequence of the `Principle
of minimal action.'

In particular, in Section \ref{Sect:Model} 
we introduce the main formulas for the kinematics,
the dynamical equations and conservation laws for the case of a
general, central self-interaction potential, and
state the main lines for the analysis of the stationary points of the
action describing the classical motion of the particle
in three spatial dimensions.
In Section \ref{Sect:HardSphere} we turn to the more specific case of a
`hard-sphere' self-interaction potential (effectively treating the
particle as a `billiard ball'), and we separately analyze the
classical motions for the cases without and with collisions.
We show (explicit and detailed formulas for the case of a coplanar
motion of the particle with respect to the wormhole's mouths
are presented in the appendix A.2)
that the action is minimized along all these trajectories, and
therefore conclude that the globally self-consistent solutions
for our model are a direct consequence of the principle of minimal
action.
We conclude in Section \ref{Sect:Discussion} with some remarks on
the reformulation of the model in terms of a Cauchy initial problem
and compare with results presented in previous literature.

\section{The model}
\label{Sect:Model}

We consider the motion of a self-interacting particle of mass $m$
in the background with a wormhole `time machine'.
The mouths of the wormhole are assumed to have a size which is
much smaller than any other scale present in the model, so that they
can be treated as pointlike, and to be infinitely heavy, so that we
can neglect the recoil effect on the geometry when the particle
traverses the wormhole.
In particular, we suppose that the mouths are at rest in some reference
frame, and consider the problem using this frame.
Moreover, spacetime outside the `time machine' is approximated to
be Minkowskian, and the motion of the particle to be non-relativistic.
We further restrict our analysis to motions in which the particle
traverses the wormhole only once.
Our discussion will be essentially independent of other features
defining the internal structure of the wormhole (although it is
consistent, e.g., with the choice of `traversal rules' suggested
in Ref.~\cite{ekt}).

The motion can be schematically described in the following way.
The particle is assumed to start at time $t_i$ in the position $\ri$,
to enter the first mouth (B) of the wormhole at time $\tb+\t$
(position $\rb$), to exit from the other mouth (A) at the earlier time $\tb$
(position $\ra$) and to finally end
its trajectory at time $t_f$ in the position $\rf$.
For the particle itself (in its proper time), the motion through the
wormhole happens almost istantaneously, as the path length of the
wormhole handle is assumed to be infinitely short.
According to an external observer, instead, the particle traversing
the time machine travels back in time by the amount $\Delta t=-\t$,
where by definition $\t>0$.

We consider the case\footnote{It is also possible to consider other
relations among the times $t_i, t_f, \bar{t}$, and $\tau$, 
but for simplicity
we shall not do this here.}
\begin{equation}
\left. \begin{array}{l}
\tb >t_i \\ 
 t_f >\bar{t} +\tau . 
       \end{array}
\right.
\label{tre}
\end{equation}

We first analyze the generic motion of the particle
in three spatial dimensions, and
then discuss the detailed trajectories for the case of
coplanar motion with respect to the wormhole's mouths.

Between times $\tb$ and $\tb+\t$ there are two copies of the same
particle, which are treated as independent objects
subject to an interaction potential $V$ of central type.
The motion can be then divided into three main regions:\par 
I) $t_i<t<\tb$~: only the first copy of the particle 
with position $\runo(t)$
is present;\par
II) $\tb<t<\tb+\t$~: both copies of the particle with positions 
$\runo(t)$ and 
$\rdue(t)$ are present;\par 
III) $\tb+\t<t<t_f$~: only the second copy of the particle with
position
$\rdue(t)$ is present.

We assume the initial and final position of the particle 
\begin{equation}
\left. \begin{array}{l}
\runo(t_i)\equiv \ri \\ 
\rdue(t_f)\equiv \rf
       \end{array}
\right.
\label{010}
\end{equation}
to be fixed. 

The time delay $\t$ in the wormhole, as well as the positions
of the wormhole's mouths
are also assumed to be known, and we neglect the interaction
between the particle and the mouths.
The entrance and exit conditions on the position of the particle are
formally summarized as the constraints
\begin{equation}
\left. \begin{array}{l}
\runo(\tb+\t)=\rb \\ 
\rdue(\tb)=\ra. 
       \end{array}
\right.
\label{011}
\end{equation}

The total action describing such a motion
is the sum of the actions of the single paths in each separate
region (subject to obvious continuity conditions for the position
of the copies of the particle at times $\tb$ and $\tb+\t$), i.e., 
\begin{eqnarray}
S&=&{m\over 2}\int^{\tb}_{t_i}dt~\druno^2(t)+{m\over
2}\int^{t_f}_{\tb+\t}dt~\drdue^2(t)+\int^{\tb+\t}_{\tb}dt\biggl\{{m\over 2}
\druno^2(t)+{m\over 2}\drdue^2(t)
\non \\
& & {} - V(\v\runo(t)-\rdue(t)\v)\biggr\}
\equiv S_1(t_i, \tb)+S_2(\tb+\t, t_f)+S_{12}(\tb, \tb+\t). 
\label{0a}
\end{eqnarray}
The general procedure consists in imposing the principle of
stationarity of the action, deriving the classical equations of motion in each
of the three Regions~I, II, and III, and solving them separately.
In the variational principle we only consider continuous paths for which
the initial and final positions are held fixed and subject to
conditions \rff{011}.

\paragraph{Regions I and III.}
By the variation of the action $S_1$ of Eq.~\rff{0a} with respect to $\runo$ in
the first region and $S_2$ with respect to $\rdue$ in the third region,
we find
\begin{equation}
\left. \begin{array}{lcl}
{\d S_1\over\d\runo}=0 \;\; & \Rightarrow &\;\;  m\ddruno=0, 
\\[0.5em] 
{\d S_2\over\d\rdue}=0 \;\; & \Rightarrow & \;\;  m\ddrdue=0. 
       \end{array}
\right.
\label{01}
\end{equation}
Equations \rff{01} clearly represent linear motion.

\paragraph{Region II.}
In this region the motion is more involved, as the two copies of
the particle interact via the potential $V$.

Let us consider the case of a general potential first.
By varying the action $S_{12}$ with respect to $\runo$ and $\rdue$
we have the following equations of motion
\begin{equation}
\left. \begin{array}{lcl}
{\d S_{12}\over\d\runo}=0\;\; & \Rightarrow & \;\; 
m\ddruno=V^{\p}(\v \rdue-\runo\v)
{(\rdue-\runo)\over \v\rdue-\runo\v}, 
\\[0.5em] 
{\d S_{12}\over\d\rdue}=0 \;\; & \Rightarrow& \;\; 
m\ddrdue=-V^{\p}(\v \rdue-\runo\v)
{(\rdue-\runo)\over \v\rdue-\runo\v}. 
       \end{array}
\right.
\label{03}
\end{equation}
For a general, position-dependent potential $V$, it is obvious from 
Eqs.~\rff{03} that the motion will no longer be linear.

We can simplify the analysis by introducing the two variables
\begin{equation}
\left. \begin{array}{l}
\r(t)\equiv \rdue(t)-\runo(t)
\\[0.5em] 
\R(t)\equiv {1\over 2}[\runo(t)+\rdue(t)]
       \end{array}
\right. 
\label{0c}
\end{equation}
Summing and subtracting Eqs.~\rff{03} we therefore have
\begin{equation}
\left. \begin{array}{l}
m\ddR(t)=0, 
\\
m\ddr(t)=-2V^{\p}[r(t)]{\r\over r}, 
       \end{array}
\right. 
\label{022}
\end{equation}
where we have defined $r\equiv |\vec{r}|$.
The new equations \rff{022} admit the integrals
\begin{equation}
\left. \begin{array}{l}
\u =\dR
\\
\E={1\over 2} \, m \, \dr^2(t)+ 2 \, V[r(t)]
       \end{array}
\right.
\label{028}
\end{equation}
where $\u = {\mbox{\em const}}$ and $\E = {\mbox{\em const}}$.
Equations~\rff{028} are nothing but the laws of energy
and momentum conservation for the motion of the two copies of the
particle in Region~II.
We incidentally note that the same integrals could have been derived
in a standard way via the variational principle, by simply noting that
the action \rff{0a} is invariant with respect to translations in both
time and space directions.

For a generic, long-range potential $V$, the trajectories will therefore
depend on the parameters\footnote{For a generic motion in three
spatial dimensions, the number of parameters
is seven, i.e.\ $\tb$ plus the three components of $\runoo$ and $\rdueo$.}
\begin{equation} 
\tb,\ \ \runo(\tb)\equiv\runoo, \ \ \rdue(\tb+\t)\equiv\rdueo. 
\label{012}
\end{equation} 
The general method is then
to verify whether there is a value (at least one)\ of
such parameters for which the total action \rff{0a} evaluated along
such trajectories has a minimum (at least local).
This would imply that the only possible
classical trajectories for the particle for which the action is
stationary and minimized are those which are globally
self-consistent.
It is in this sense that the `Principle of self-consistency' [7, 9,
13--15],
can be looked at as
a direct consequence of the `Principle of minimal action.'

In general, the problem of solving the equations of motion and
minimizing the classical action for a generic potential $V$ is not
straightforward.
In the case, e.g., of a Coulomb-like repulsive interaction between the
two copies of the particle, the problem can be shown to finally reduce to
that of looking for the
stationary points of the classical action with respect to
the parameters (eccentricity, semimajor axis etc.)\ of the hyperbolic
orbit for $\r$.
The system of equations involved in the procedure, however, cannot be
solved exactly in an analytic form, but only by making some specific ansatz
and using, e.g., some perturbative expansion.
We hope to turn back to this case in a future work.

\section{The case of a `hard-sphere' potential}
\label{Sect:HardSphere}

The nature of the trajectories can be greatly simplified, instead,
by working with a `short-range' potential.
We can assume, for instance, that the two copies of the particle are
like two (small)\ `billiard balls,' interacting via a hard-sphere
potential of the kind
\begin{equation} 
V(r)=V_1\theta (r_s-r);\ \ 
V_1\rightarrow\infty,\ \
r_s\rightarrow
0. 
\label{2}
\end{equation} 
In this approximation, we essentially neglect the interaction of the
two copies of the particle along most of 
their motion in Region~II, except at
the point of the (eventual) collision,
which we assume to be essentially elastic
(we also neglect the deformation of the `balls').
The effect of the potential is limited to an infinitesimally small period
of time around $\to$, the time of the (eventual) collision.

For such a potential, it is clear that the second of Eqs.~\rff{022} and
\rff{028} are not even well defined at the point $r=r_s$.
However, also in this case it is still possible to show 
(see Appendix~\ref{App:Hard})\ 
that the kinetic energy $\dr^{~2}$ is conserved before and after
the collision.

The total momentum is also conserved during the whole motion in
Region~II (and also before and after the eventual collision point), as clearly
the first of Eqs.~\rff{028} shows.

Finally, taking the variation of the action \rff{0a} with respect to
$\tb$, and excluding the possibility of collisions on the verge of the
wormhole's mouths (in other words, if we assume that, for the short-range
potential \rff{2}, $V[r(\tb)]=V[r(\tb+t)]=0$), we get the condition
\begin{equation} 
[\druno(\tb+\t)]^2=[\drdue(\tb)]^2
\lb{wor}
\end{equation} 
stating that the energy of the particle at its entrance and exit from the
wormhole's mouths must be conserved.
The condition \rff{wor} had also been identified by the authors of 
Ref.~\cite{ekt} 
as one of the wormhole `traversal rules'.

Moreover, for $r>r_s\sim 0$ we have  $V(r)=0$, and
Eqs.~\rff{03} are now perfectly well
defined: they state that, everywhere in Region~II except at the point
of eventual collision, the motion of the two copies of the particle
is also linear.

In conclusion, for fixed initial and final positions
of a particle constrained to traverse the wormhole once,
we have to distinguish between the two cases:

\paragraph{i) trajectories without self-collision.}
In this case, the first copy of the particle moves linearly from the
initial position $\ri$ at time $t_i$ until it enters the wormhole
mouth B at time $t=\tb+\t$.
Similarly, the second copy of the particle moves linearly
from the mouth A at time $\tb$ up to the final position $\rf$ at time
$t_f$.

\paragraph{ii) trajectories with self-collision.}
In this case, instead, the motion for the first (second)\
 copy of the particle
is linear from the initial position $\ri$ at time $t_i$
(from the wormhole mouth A at time $\tb$) up to the collision event, with
coordinates
\begin{equation} 
\runo(\to)=\rdue(\to)\equiv\ro. 
\lb{311b}
\end{equation} 
After the collision, the motion of the first (second)\ copy of the particle
is linear again up to the wormhole mouth B at time $\tb +\t$
(up to the final position $\rf$ at time $t_f$).
Of course, the directions of the motions for the first (second)\ copy
of the particle will be, in general, different before and after the
collision (see Section~\ref{Sect:Sub:WSC}).

In the case of a short-range potential, therefore, the stationarity
problem is easier than for the general potential case.
The trajectories will depend, in fact, only on the parameters
\footnote{In three spatial dimensions, these are five `degrees of freedom'
instead of the seven of the general case.}
\begin{equation} 
\tb,\ \ \to,\ \ \ro. 
\label{012new}
\end{equation} 
The problem is now to look for the stationary points (if any)\ 
of the action
\rff{0a}, evaluated along the classical 
trajectories \rff{01}, \rff{03}, 
and \rff{022}, with respect to the 
parameters \rff{012new}, and to see if
they are minima.
We first consider in details the variational problem for the
case of no collision (i.e., when the separation $r$ between the
two copies of the particle is always greater than $r_s$), so as
to show how the main lines of the analysis proceed.

\subsection{Trajectories without self-collision}
\label{Sect:Sub:NSC} 

In the no-collision case, the solutions
of the equations of motion \rff{01}--\rff{03},
subject to the boundary conditions \rff{010}--\rff{011}, are given
by the linear trajectories
\begin{equation}
\left. \begin{array}{l}
\runo(t; \tb)={[\rb(t-t_i)+\ri(\tb+\t-t)]\over (\tb+\t-t_i)}, 
\\[0.5em] 
\rdue(t; \tb)={[\ra(t_f-t)+\rf(t-\tb)]\over (t_f-\tb)}. 
       \end{array}
\right.
\lb{387}
\end{equation}
The variational problem is now extremely simple, as the action
\rff{0a} evaluated along the classical trajectories \rff{387}
is a function only of the parameter $\tb$.
Noting that for these classical
trajectories, the contribution of the potential term to the
action \rff{0a} is identically zero (see Appendix~\ref{App:Hard}),
and using Eqs.~\rff{387} for the kinetic terms, the classical action
for the no-collision case ($S_{cl, n-c}$) becomes
\begin{equation}
S_{cl, n-c}(\tb)={m\over 2}\left [{(\rb-\ri)^2\over (\tb+\t-t_i)}+
{(\rf-\ra)^2\over (t_f-\tb)}\right]. 
\lb{388}
\end{equation}
It is now easy to see that variation of Eq.~\rff{388} with respect
to $\tb$ gives again condition \rff{wor} for the case of the
trajectories \rff{387}.
Solving Eq.~\rff{wor} for $\tb$ we get
\begin{equation}
\tb={[t_f\pm(t_i-\t)W]\over (1\pm W)}
\lb{286}
\end{equation}
where we have denoted $W\equiv\v\rf-\ra\v/\v\rb-\ri\v$.

We have to check that in finding the two stationary solutions
\rff{286} for $\tb$ we have not inadvertently violated the conditions
\rff{tre}, i.e. that
\begin{equation}
t_i<\tb<t_f-\tau. 
\lb{63}
\end{equation}
It is easy to see that the only solution satisfying the condition
\rff{63} is given by Eq.~\rff{286} with the plus sign, subject to the
constraint that
\begin{equation}
t_f-t_i > \max\left\{ W\tau;\ \  W^{-1}\tau \right\}. 
\lb{284}
\end{equation}
Substituting Eq.~\rff{286} back into Eqs.~\rff{387} finally gives
the explicit trajectories
\begin{equation}
\left. \begin{array}{l}
\runo(t)=\ri+(1+W)(\rb-\ri){(t-t_i)\over (t_f-t_i+\t)}, 
\\[0.5em] 
\rdue(t)=\rf+(1+W^{-1})(\rf-\ra){(t-t_f)\over (t_f-t_i+\t)}. 
      \end{array}
\right.
\lb{294}
\end{equation}
The stationary point \rff{286} is also clearly a minimum of the action
$S_{cl, n-c}$, since
\begin{equation}
{d^2 S_{cl, n-c}\over d\tb^2}\biggr\v_{stat}= m
\left [{(\rb-\ri)^2\over (\tb+\t-t_i)^3}+{(\rf-\ra)^2\over (t_f-\tb)^3}\right ]
>0. 
\lb{393}
\end{equation}
We conclude that the problem of a `billiard ball'-like particle moving
between specified initial and final positions, subject to the conditions
of traversing the wormhole once and having no self-collisions,
has a unique globally
self-consistent solution which can be derived by simply imposing
the principle of minimal action.

\subsection{Trajectories with self-collision}
\label{Sect:Sub:WSC} 

In the case of self-collision under the action of the `hard-sphere'
potential \rff{2}, as we already remarked at the beginning of 
Section \ref{Sect:Model},
the motion is also linear everywhere except at the event of the impact.
To clearly identify the trajectories before and after the collision,
it is convenient to slightly modify the notation defining
\begin{equation}
\left. \begin{array}{ll}
\runo(t),\ \ &t_i<t<\to, 
\\
\runop(t),\ \ & \to<t<\tb+\tau,  
       \end{array}
\right.
\lb{newnot1}
\end{equation}
for the first copy of the particle, and
\begin{equation}
\left. \begin{array}{ll}
\rdue(t),\ \ &\tb<t<\to, 
\\
\rduep(t), &\to<t<t_f, 
       \end{array}
\right.
\lb{newnot2}
\end{equation}
for the second copy of the particle.

It is then easy to show that the solutions of the classical equations of motion
\rff{01}--\rff{03}, subject to the boundary conditions
\rff{010}--\rff{011}, and \rff{311b}, are given by
\begin{equation}
\left. \begin{array}{l}
\runo(t; \tb, \to, \ro)={[\ro(t-t_i)+\ri(\to -t)]\over \toi}, 
\\[0.5em] 
\rdue(t; \tb, \to, \ro)={[\ro(t-\tb)+\ra(\to -t)]\over \tob}, 
\\[0.5em] 
\runop(t; \tb, \to, \ro)={[\rb(t-t_0)+\ro(\tb+\t -t)]\over \tbto},  
\\[0.5em] 
\rduep(t; \tb, \to, \ro)={[\rf(t-t_0)+\ro(t_f -t)]\over \tfo}. 
       \end{array}
\right. 
\label{314}
\end{equation}
With the new notation \rff{newnot1}--\rff{newnot2}, the action \rff{0a}
more simply reads
\begin{eqnarray}
\hat S&=&{m\over 2}\biggl[\int^{\to}_{t_i}dt~[\druno(t)]^2+\int^{\tb+\t}_{\to}
dt~[\druno^{\prime}(t)]^2+\int^{\to}_{\tb}dt~[\drdue(t)]^2
\non \\
& & {} + \int^{t_f}_{\to}
dt~[\drdue^{\prime}(t)]^2\biggr ]-\int_{\tb}^{\tb+\t}dt~V(\v\runo(t)-
\rdue(t)\v)
\label{310}
\end{eqnarray}
In order to evaluate the total action $\hat S_{cl, c}$ for the collision case
along the classical trajectories described by Eqs.~\rff{314},
we can use arguments similar to those in Section~\ref{Sect:Sub:NSC} 
(see Appendix~\ref{App:Hard})\ to show
that the contribution of the potential term is again zero.

Therefore, using the classical trajectories \rff{314} for the
kinetic terms in Eq.~\rff{310}, we obtain
\begin{equation}
\hat S_{cl, c}(\tb, \to, \ro)
={m\over 2}\left [{\roi^2\over \toi}+{\rob^2\over \tbto}+
{\roa^2\over \tob}+{\rof^2\over \tfo}\right]. 
\lb{315}
\end{equation}
The variational problem for the collision case consists in looking for
the stationary points of the action \rff{315} with respect to the
parameters $\ro$, $\to$, and $\tb$.
Taking derivatives with respect to $\ro$, $\to$, and $\tb$, 
and defining
the velocities $\vuno\equiv\druno$, $\vdue\equiv\drdue$, 
$\vunop\equiv\drunop$, 
and $\vduep\equiv\drduep$, from Eqs.~\rff{314} and \rff{315}
we find the following conditions
\begin{equation}
\left. \begin{array}{l}
\vuno+\vdue =\vunop+\vduep, 
\\[0.5em] 
(\vuno)^2+(\vdue)^2 =(\vunop)^2+(\vduep)^2, 
\\[0.5em] 
(\vunop)^2 =(\vdue)^2. 
       \end{array}
\right.
\lb{319}
\end{equation}
These equations respectively represent the conservation laws for
momentum and  energy during the collision, and the conservation of
energy at the entrance and exit of the particle at the two wormhole's
mouths (cf.\ Ref.~\cite{ekt}).

Equations \rff{319} can in principle 
be solved either directly in the $\ro$, $\to$, and $\tb$
variables, or in terms of the velocity variables (for instance,
considering $\vunop$ and $\vduep$ as unknowns and $\vuno$ and $\vdue$ as
parameters, the number of equations and unknowns in the problem
remaining the same). 

Using velocities as our unknowns and introducing the quantities
\begin{equation}
\left. \begin{array}{l} 
\vec{a}\equiv\vuno-\vduep =\vunop-\vdue, 
\\[0.5em] 
\vec{b}\equiv\vunop+\vdue, 
\\[0.5em] 
\vec{c}\equiv\vuno+\vduep, 
       \end{array}
\right.
\lb{319a}
\end{equation}
Eqs.~\rff{319} can be easily transformed into the equivalent system
of conditions
\begin{equation}
\left. \begin{array}{l}
\vec{a}\cdot\vec{b}=0,
\\[0.5em]
\vec{a}\cdot\vec{c}=0.
       \end{array} 
\right.
\lb{319b}
\end{equation}
For the motion in three spatial dimensions, the most general solution
of the conservation laws \rff{319} is thus given by
\begin{equation}
\left. \begin{array}{l}
\vuno={1\over 2}(\vec{c}+\vec{a}), 
\\[0.5em] 
\vdue={1\over 2}(\vec{b}-\vec{a}), 
\\[0.5em] 
\vunop={1\over 2}(\vec{b}+\vec{a}), 
\\[0.5em] 
\vduep={1\over 2}(\vec{c}-\vec{a}), 
      \end{array}
\right. 
\lb{319c}
\end{equation}
for any {\em arbitrary\/} $\vec{a}$ which is {\em orthogonal\/} to 
{\em arbitrary\/} $\vec{b}$ and $\vec{c}$.
Then, in principle, in the case of a generic three dimensional 
motion Eqs.~\rff{319b} (with $\vec{a}$ given by the first of 
Eqs.~\rff{319a})\ can be solved, 
using Eqs.~\rff{314}, for $\ro$, $\to$, and $\tb$ 
and therefore for the complete trajectories.
We shall not do that here, but only consider the simpler case of two
dimensional spatial motion.

\subsubsection{Coplanar motion}

The solutions of the conservation Eqs.~\rff{319} for the case of
two dimensional, coplanar motion of the copies of the particle with
respect to the wormhole's mouths can be deduced from the generic three
dimensional solutions \rff{319c} by restricting to the following
ans\"atze for $\vec{a}, \vec{b}$ and $\vec{c}$
\begin{equation}
{\mbox{\em i)}}\ \ \vec{a}=\vec{0};\ \ {\mbox{or}}\ \
{\mbox{\em ii)}}\ \ \vec{c}=\epsilon\vec{b}. 
\lb{319d}
\end{equation}
where $\epsilon$ is an arbitrary constant.

In particular, the ansatz 
{\em i)}\ 
of 
Eq.~\rff{319d} 
corresponds to the case of

\paragraph{a) `Velocity exchange' rule:}
\begin{equation}
\left. \begin{array}{l} 
\vunop=\vdue,
\\[0.5em] 
\vduep=\vuno,
       \end{array}
\right.  
\label{19}
\end{equation}
while the ansatz {\em ii)\/} of Eq.~\rff{319d} (for $\epsilon \neq -1$)
corresponds to the case of

\paragraph{b) `Mirror exchange of velocities' rule:}
\begin{equation}
\left. \begin{array}{l} 
(v_1^{\prime})_x=(v_2)_x
\\[0.5em] 
(v_2^{\prime})_x=(v_1)_x
\\[0.5em] 
(v_1^{\prime})_y=-(v_2)_y
\\[0.5em] 
(v_2^{\prime})_y=-(v_1)_y
       \end{array}
\right.
\label{19i}
\end{equation}
(expressed in component notation, where we have chosen the $x$-axis to
lie along the direction of $\vuno+\vdue =\vunop+\vduep$).

For the particular value $\epsilon=-1$, the ansatz 
{\em ii)\/} of Eq.~\rff{319d} 
no longer corresponds to the `topology' of the `mirror exchange
solutions,' but to the case of

\paragraph{c) `Collinear velocities' rule:}$\!\!\!\!$\footnotemark 
\footnotetext{This case apparently was not considered in Ref.~\cite{ekt}.}
\begin{equation}
\left. \begin{array}{l}
\vuno =-\vdue
\\[0.5em] 
\vunop =-\vduep
\\[0.5em] 
\v\vuno\v=\v\vdue\v~=~\v\vunop\v~=~\v\vduep\v. 
       \end{array}
\right.
\lb{320}
\end{equation}
These solutions are `degenerate' in the sense that the velocities
$\vuno$ and $\vdue$ must be along the direction identified by $\ri$ and
$\ra$ (i.e. the velocity of the first copy of the particle must
be initially pointing towards the wormhole mouth A), and similarly
the velocities $\vunop$ and $\vduep$ must be along the direction
identified by $\rf$ and $\rb$ (i.e. the velocity of the second copy of
the particle after the collision must be outwards pointing from
the wormhole mouth B).\footnote{The solution $\vec{c}=-\vec{b},
\vec{a}=\vec{0}$ is a `doubly degenerate' case, as it corresponds
to one dimensional spatial motion of the two copies of the particle
along the line connecting the wormhole's mouths.}
It is also easy to show that the velocity rules \rff{320}, combined with
the condition that the total duration of the motion (as seen from an
observer external and at rest with respect to the wormhole) is fixed
and equal to $t_f-t_i+\t$, limit the possible choices of the boundary
data $\ri, \rf, t_i, t_f$.
In particular, only five of such data (e.g., $\ri, t_i$ and $\rf$)
can be arbitrarily chosen, while the sixth (e.g. $t_f$) will be constrained.

Finally, there is also the `trivial' solution in which the
velocities of the copies of the particle do not change before and
after the collision.
This actually corresponds to the no-collision case which we already considered
in Section \ref{Sect:Sub:NSC}. 
\subsubsection{Stationary points for the coplanar motion}
Let us consider
the three nontrivial solutions of the conservation Eqs.~\rff{319}.

As they stand, solutions \rff{19}--\rff{320} are still implicit equations
for the original variables $\ro, \to$ and $\tb$, which are the final
object of our analysis of the stationary point of the action \rff{0a}.
The algebra leading to
the `solutions' for the velocity rules {\em a--c\/} in terms of
the collision coordinates and $\tb$ is not particularly interesting and
we leave the details for the interested reader in the 
Appendix~\ref{App:Traj}.

The main result of such an analysis is that, for each of the cases
{\em a--c,} there exists a 
unique solution for which the action \rff{315} is
stationary and minimized with respect to the 
parameters $\ro$,
$\to$, and $\tb$ (the stationary values of $\to$, $\tb$, and $\ro$
in each of the three cases are respectively 
given by Eqs.~\rff{324}--\rff{326}
for case {\em a,} Eqs.~\rff{353}, \rff{144}, and
\rff{350} for case {\em b,} and finally Eqs.~\rff{362}--\rff{363}
for case {\em c}).
In other words, if we fix the initial and final conditions \rff{010}
and the boundary conditions \rff{011},
the variational problem for the action \rff{0a}, subject to the condition
of self-collision for the particle,
admits only the set of globally self-consistent classical trajectories
given by Eqs.~\rff{296}--\rff{301}, \rff{mirtraj}, \rff{368},
each of which is subject, respectively, to the constraints \rff{177},
\rff{145} and \rff{355}, \rff{new} and \rff{366}.

We have thus proved that,
for the model of a particle which is constrained to traverse a wormhole
time-machine geometry once and to self-interact via the
`hard-sphere' potential \rff{2},  the whole set of
classical trajectories which are
globally self-consistent can be directly and simply recovered
by imposing the principle of minimal action.

\section{Discussion}
\label{Sect:Discussion}

If the laws of physics actually permit the existence of traversible
wormholes, then in principle it would be
possible to convert the wormhole into a time machine and therefore
have CTCs looping through it.

The analysis of the present paper is purely classical.
We have considered the simple model of a nonrelativistic
particle which is constrained
to have fixed initial and final positions, to loop through the wormhole
once and to interact with itself by means of a `hard-sphere', elastic
potential $V$.
We used the action principle to derive the classical trajectories, and
we found that the only possible solutions which minimize the action
are those which are globally self-consistent.
In the case of coplanar motion with respect to the wormhole's mouths,
the possible, globally self-consistent trajectories in which the
particle's copies collide are of three types, depending on the three
possible ways in which momentum is exchanged (i.e., `velocity',
`mirror' or `collinear' exchange) at the collision event.
In this way we have shown that the `Principle of self-consistency'
can be 
actually 
encoded in a natural way into the `Principle of
minimal action'.

In a previous series of papers \cite{ekt,nov5},
the analogous model for the
motion of a nonrelativistic `billiard ball'-like particle moving
in the spacetime containing a wormhole time machine has been
considered in the context of a Cauchy initial value problem (for
a general discussion also including the case of a scalar field, see
Ref.~\cite{con}).
In the case of elastic, no-frictional self-interaction between the
two copies of the particle, it has been shown \cite{ekt} that generic classes
of initial data have multiple, and even infinite numbers of globally
self-consistent solutions to the equations of motion
(trajectories where the particle is initially at rest far from the
wormhole have, in fact, multiplicity one), with no evidence for
non self-consistent trajectories.
The extension to the case of an inelastic self-collision of
`billiard ball'-like particles was made in Ref.~\cite{nov5}.

Our globally self-consistent solutions for the case of collision
with `velocity exchange' and `mirror exchange of velocities'
are in fact consistent with the solutions found in Ref.~\cite{ekt}.
A more detailed analysis of the multiplicity of solutions and
the connection between the Cauchy and the boundary value problems,
together with a consideration of the back-reaction effects of the
particle motion on the wormhole geometry, and of more complicated
self-interaction potentials for the particle itself,
will be made in a future publication.

In this paper we have not addressed the analysis of the conditions under
which a wormhole can be created and maintained in a Lorentzian
spacetime (see also Ref. \cite{vissb}).
It is known that static wormholes require a matter which violates
the averaged weak energy condition (AWEC) [3, 5]
(and if spacetime has to maintain a well defined spinorial structure on
it, then the creation of wormholes should also occur in pairs
\cite{gibb}).
Classically, this condition might be a problem, since ordinary classical energy
densities are positive.\footnote{A new interesting set of dynamical wormhole
solutions has been recently discovered \cite{dyn} which satisfy the weak and
dominant energy conditions, although it is not clear that these
wormholes can be traversed.}

On the other hand, it is not clear whether quantum effects
could anyway preserve the AWEC for generic cases (a well known
counterexample is the Casimir effect).
Whether vacuum polarization effects in the quantum theory
can actually destabilize a wormhole (due to the infinite back-reaction
caused by a diverging renormalized stress-energy tensor)
is still a challenging issue (for two approaches with opposite
conclusions see, e.g., Refs.~\cite{qua1,qua2}).

The results of this paper motivate us to formulate the conjecture that
the `Principle of self-consistency' is a consequence of the `Principle
of minimal action' in the general case for all physical phenomena,
and not only for the simple mechanical problem considered here.

\vspace{33pt}

\noindent {\Large \bf Acknowledgements}{\vspace{11pt}}

This work is supported in part by Danish Natural Science Research
Council through grant N9401635 and in part by Danmarks
Grundforskningsfond through its support for the establishment of the
Theoretical Astrophysics Center.
A.C.'s research is supported by an EEC fellowship in the `Human Capital
and Mobility' program, under contract No. ERBCHBICT930313.

\newpage

\appendix
\section{Appendix}
\subsection{The `hard-sphere' potential}
\label{App:Hard}

In this appendix we prove some interesting properties for the
`hard-sphere' potential of Eq.~\rff{2}.\footnote{These results are valid
for the generic case of motion in three spatial dimensions.}

$\bullet$ {\it Energy conservation}

For the potential \rff{2} the second of conservation Eqs.~\rff{028} is
apparently ill defined at the collision event.
What, however,  this equation clearly says is that,
in the region $r<r_s$, the motion is not classically allowed, as
the kinetic energy for $\r$ would be infinitely large and negative.
Assuming that the classical motion proceeds until $r=r_s$ (i.e., there
is the collision at time $\to$), we can now integrate both sides
of the second of
Eqs.~\rff{022} around $r_s$, obtaining
\begin{eqnarray}
{I\over m}~&\dot =&\int_{r_s+\epsilon_1}^{r_s}d\r \cdot \ddr +
\int^{r_s+\epsilon_2}_{r_s}d\r \cdot \ddr
\non \\
&=&\int^{\to}_{\to-\delta t_1}dt~{d\over dt}(\dr)^2
+\int_{\to}^{\to+\delta t_2}dt~{d\over dt}(\dr)^2=
[\dr(\to+\delta t_2)]^2-[\dr(\to -\delta t_1)]^2
\lb{004}
\end{eqnarray}
(with $\epsilon_1, \epsilon_2, \delta t_1, \delta t_2>0$)
when acting on the left hand side, while
\begin{equation}
I= 2V_1\int_{r_s+\epsilon_1}^{r_s+\epsilon_2}d\r \cdot \left
[{\r \over r}~\delta(r-r_s)\right ]=2V_1\int_{r_s+\epsilon_1}^{r_s+\epsilon_2}
dr~ \delta(r-r_s)=0
\lb{004bis}
\end{equation}
when acting on the right hand side.
{}From Eqs.~\rff{004} and \rff{004bis} we conclude that the kinetic
energy $\dr^{~2}$ is conserved before and after the collision, as expected.

$\bullet$ {\it Zero contribution to classical action}

Let us consider the contribution of the potential \rff{2} to the action
\rff{0a} evaluated along the classical trajectories \rff{387} or
\rff{314}.

In the no-collision case, denoting with $t_m$ and $r_m$ the time
and position of minimum distance
between the two copies of the particle, we have, for the potential
\rff{2}
\begin{eqnarray}
\int_{\tb}^{\tb+\t}dt~V(r)\biggr\v_{cl} &=&
V_1\left [\int_{\tb}^{t_m}dt~\theta(r_s-r)
+\int_{t_m}^{\tb+\t}dt~\theta(r_s-r)\right ]
\non \\
&=& V_1\left [\int_{r(\tb)}^{r_m}{dr\over \dot r}~\theta(r_s-r)
+\int_{r_m}^{r(\tb+\t)}{dr\over \dot r}~\theta(r_s-r)\right ] =0
\lb{005}
\end{eqnarray}
since obviously, in the no-collision case, $r_m, r(\tb)$ and $r(\tb+\t)$ are
always greater than $r_s$.

Similarly, in the collision case, in the limit $r_s\rightarrow 0$,
the motion is such that (except
at the point of collision) the angular part ($\dot \theta$)
of the velocity $\dr$ is essentially zero, and
using this result together with the conservation equation for $\dr^2$
gives again
\begin{eqnarray}
\int_{\tb}^{\tb+\t}dt~V(r)\biggr\v_{cl} &=&
V_1\left [\int_{\tb}^{\to}dt~\theta(r_s-r)
+\int_{\to}^{\tb+\t}dt~\theta(r_s-r)\right ]
\non \\
&\simeq& {V_1\over \dot r}\left [\int_{r(\tb)}^{r_s}dr~\theta(r_s-r)
+\int_{r_s}^{r(\tb+\t)}dr~\theta(r_s-r)\right ] =0
\lb{005x}
\end{eqnarray}

In conclusion, the contribution of the `hard-sphere' potential $V$
to the classical action can always be neglected both in the collision
and no-collision cases.

\subsection{Trajectories for the coplanar motion}
\label{App:Traj}

$\bullet$ {\it `Velocity exchange'}

For the case of `velocity exchange' between the two copies of the
particle at the collision event, the solutions \rff{19} can be
rewritten, using Eqs.~\rff{314}, in terms of the variables $\ro, \to$
and $\tb$ as
\begin{eqnarray}
0&=&{\roi\over \toi}+{\rof\over \tfo}
\non \\
0&=&{\roa\over \tob}+{\rob\over \tbto}
\lb{321}
\end{eqnarray}
We can solve this system by first writing, from the first of 
Eqs.~\rff{321}, $\ro$ as a function of $\to$ and then inserting
this result into the second of Eqs.~\rff{321}.
We thus obtain
a set of two equations (one for each component of the vectors
$\r_{\alpha}$, with $\alpha = 0, i, f, A, B$)
which can be solved in terms of $\tb$ and
$\to$, giving
\begin{equation}
\to={\{[(\rb-\ra)\we(t_f\ri-t_i\rf)]_3+[\ra\we\rb]_3\tfi\}\over
[(\rf-\ri)\we(\rb-\ra)]_3 }
\lb{324}
\end{equation}
and
\begin{eqnarray}
\tb &=&\{[(\rb-\ra)\we(t_f\ri-t_i\rf)]_3+[\ra\we\rb]_3\tfi
\non \\
&+&[\ri\we\rf+(\rf-\ri)\we\ra)]_3\t\}\cdot\{[(\rf-\ri)\we(\rb-\ra)]_3\}^{-1}
\lb{47b}
\end{eqnarray}
where we have `artificially' defined vectors in three-dimensional space
(i.e., we define $\r_{\alpha}\dot =(x_{\alpha}, y_{\alpha}, 0)$),
and the notation $]_3$ means the third component of the vector.

Substituting these results into Eq.~\rff{321} then also explicitly gives the
spatial coordinates of the collision event, i.e.
\begin{equation}
\ro={\{(\rf-\ri)[\ra\we\rb]_3+\ri[\rf\we(\rb-\ra)]_3 -\rf[\ri\we(\rb-\ra)]_3\}
\over [(\rf-\ri)\we(\rb-\ra)]_3}
\lb{326}
\end{equation}

Finally, we can insert the formulas \rff{324}-\rff{326}
into Eqs.~\rff{314} to find out the classical trajectories as
\begin{equation}
\runo(t)=\rduep(t)={[\ri(t_f-t)+\rf(t-t_i)]\over \tfi}
\lb{296}
\end{equation}
and
\begin{eqnarray}
\runop(t)&=&\rdue(t)=\ra+\{(\ra-\rb)[(\rb-\ra)\we(t_f\ri-t_i\rf)]_3
\non \\
&+&[\ra\we\rb]_3\tfi+[\ri\we\rf+(\rf-\ri)\we\ra]_3\t
\non \\
&+&[(\rb-\ra)\we(\rf-\ri)]_3t\}\cdot\{\t[(\rf-\ri)
\we(\rb-\ra)]_3\}^{-1}
\lb{301}
\end{eqnarray}

Similarly as done for the case of no-collision, we also have to check that the
solutions for $\to$ and $\tb$ given by Eqs.~\rff{324}-\rff{47b}
satisfy the conditions
\begin{eqnarray}
t_i&<&\tb~<~t_f-\t
\non \\
\tb&<&\to~<~\tb+\t
\lb{327}
\end{eqnarray}
The first of conditions \rff{327} implies
\begin{eqnarray}
0&<&{\{[(\rb-\ra)\we\ri+\ra\we\rb]_3\tfi+[\ri\we\rf+(\rf-\ri)\we\ra]_3
\t\}\over [(\rf-\ri)\we(\rb-\ra)]_3}
\non \\
0&>&{\{[(\rb-\ra)\we\rf+\ra\we\rb]_3\tfi+[\ri\we\rf+(\rf-\ri)\we\rb]_3
\t\}\over [(\rf-\ri)\we(\rb-\ra)]_3}
\lb{66}
\end{eqnarray}
while the second of conditions \rff{327} implies
\begin{eqnarray}
0&>&{[\ri\we\rf+(\rf-\ri)\we\ra]_3\over [(\rf-\ri)\we(\rb-\ra)]_3}
\non \\
0&<&{[\ri\we\rf+(\rf-\ri)\we\rb]_3\over [(\rf-\ri)\we(\rb-\ra)]_3}
\lb{329}
\end{eqnarray}
Equations \rff{66} and \rff{329} can be simultaneously satisfied for
\begin{eqnarray}
0&<&[\ri\we\rf +(\rf-\ri)\we\ra]_3
\non \\
0&>&[\ri\we\rf +(\rf-\ri)\we\rb]_3
\non \\
0&>&[(\rb-\ra)\we\ri+\ra\we\rb]_3\tfi+[\ri\we\rf+(\rf-\ri)\we\ra]_3\t
\non \\
0&<&[(\rb-\ra)\we\rf+\ra\we\rb]_3\tfi+[\ri\we\rf+(\rf-\ri)\we\rb]_3\t
\lb{177}
\end{eqnarray}
or, equivalently, if all the inequality signs in \rff{177} are
reversed.
Conditions \rff{177} are thus the constraints on the classical
trajectory for the `velocity exchange' rule.

$\bullet$ {\it `Mirror exchange of velocities'}

The main lines of the analysis for the `mirror exchange of
velocities' case are essentially the same as for the `velocity exchange'
case.
Here, however, due to the rules \rff{19i}, it is more convenient to use
a somewhat more involved component notation, where
all vector quantities are expressed as $\vec{r}_{\alpha}~\dot=~(x_{\alpha},
y_{\alpha})$.

Using this notation and Eqs.~\rff{314}, the rules \rff{19i} in terms
of $\xo, \yo, \to$ and $\tb$ explicitly become
\begin{eqnarray}
0&=&{\xob\over \tbto}+{\xoa\over \tob}
\non \\
0&=&{\xof\over \tfo}+{\xoi\over \toi}
\non \\
0&=&{\yob\over \tbto}-{\yoa\over \tob}
\non \\
0&=&{\yof\over \tfo}-{\yoi\over \toi}
\non \\
0&=&{\yoi\over \toi}+{\yoa\over \tob}
\lb{332}
\end{eqnarray}
The last of Eqs.~\rff{332} is an extra  constraint which must be imposed
in the particular coordinate frame that we are using,
for which we have that $(v_1)_y+(v_2)_y=0$.

{}From the second and fourth of conditions \rff{332}, we can solve for
$\xo$ and $\yo$ as functions of $\to$,
and inserting these results into the remaining equations of
the system \rff{332} we get
three equations for the two `unknowns' $\to$ and $\tb$.
This means that the trajectory will be constrained as
in the `velocity exchange' case.

The explicit stationary solutions for $\to$ and $\tb$ are
\begin{equation}
\to=t_i+{[\xa(\yi-\yb)+\xb(\yi-\ya)+\xi(\ya+\yb-2\yi)](t_f-t_i)\over
[(\xb-\xa)(\ya-\yb)\tfi+(\xf-\xi)(2\yi-\ya-\yb)\t]}\t
\lb{353}
\end{equation}
\begin{eqnarray}
\tb &=& t_i+\t\{[\xa(\ya+\yi-2\yb)+\xb(\yi-\ya)+2\xi(\yb-\yi)]\tfi
\non \\
&+&(\xf-\xi)(\ya-\yi)\t\}\{(\xa-\xb)(\yb-\ya)\tfi
\non \\
&+& (\xf -\xi)(2\yi-\ya-\yb)\t\}^{-1}
\lb{144}
\end{eqnarray}
while the constraint is
\begin{eqnarray}
0&=&\t^2 (\xf-\xi)(\yf-\yi)+\t\tfi [(\xi+\xf)(\ya+\yb)-2(\xi\yf+\xf\yi)
\non \\
&+&(\xa+\xb)(\yi+\yf)-2(\xa\yb+\xb\ya)]+\tfi^2(\xa-\xb)(\ya-\yb)
\lb{145}
\end{eqnarray}

Using Eqs.~\rff{353}--\rff{144} back into the system \rff{332}, we
finally find the spatial coordinates of the collision event as
\begin{eqnarray}
\xo&=&\xi+{[\xa(\yi-\yb)+\xb(\yi-\ya)+\xi(\ya+\yb-2\yi)](\xf-\xi)\over
[(\xb-\xa)(\ya-\yb)\tfi+(\xf-\xi)(2\yi-\ya-\yb)\t]}\t
\non \\
\yo&=&\yi+\{[\xa(\yb-\yi)+\xb(\ya-\yi)+\xi(2\yi-\ya-\yb)](\yf-\yi)\t\}
\non \\
&\times &\{(\xb-\xa)(\ya-\yb)\tfi+[2\xa(\yb-\yi)
\non \\
&+&2\xb(\ya-\yi)+(\xf+\xi)(2\yi-\ya-\yb)]\t\}^{-1}
\lb{350}
\end{eqnarray}
and, using Eqs.~\rff{353}--\rff{144} and \rff{350}, the trajectories
\rff{314} explicitly become
\begin{eqnarray}
x_1&=&\xi+{(x_f-x_i)\over\tfi}(t-t_i)
\non \\
y_1&=&\yi+\{[(\xa-\xb)(\ya-\yb)\tfi+(\xf-\xi)(\ya+\yb-2\yi)\t]
\non \\
&\times &(\yf-\yi)(t-t_i)\}\{(\xb-\xa)(\ya-\yb)\tfi+[2\xa(\yb-\yi)
\non \\
&+&2\xb(\ya-\yi)+(\xf+\xi)(2\yi-\ya-\yb)\t]\}^{-1}
\non \\
x_2&=&\xa+(\xb-\xa)\biggl \{ {(t-t_i)\over \t}+\{(\xf-\xi)(\yi-\ya)\t
\non \\
&+&[\xa(2\yb-\ya-\yi)+\xb(\ya-\yi)+2\xi(\yi-\yb)]\tfi\}
\non \\
&\times& [(\xb-\xa)(\ya-\yb)\tfi+(\xf-\xi)(2\yi-\ya-\yb)\t]^{-1}\biggr\}
\non \\
y_2&=&\ya+\biggl\{(\xb-\xa)(\ya-\yb)(\yi-\ya)\tfi+\biggl [ [\xa(\yb-\yi)
\non \\
&+&\xb(\ya-\yi)](\yi+\yf-2\ya)+[\xf(\yi-\ya)+\xi(\yf-\ya)]
\non \\
&\times&(2\yi-\ya-\yb)\biggr ]\t\biggr \}
\biggl \{[\xa(2\yb-\ya-\yi)+\xb(\ya-\yi)
\non \\
&+&2\xi(\yi-\yb)]\tfi+(\xf-\xi)(\yi-\ya)\t
\non \\
&+&[(\xb-\xa)(\ya-\yb)\tfi+(\xf-\xi)(2\yi-\ya-\yb)\t]{(t-t_i)\over \t}
\biggr \}
\non \\
&\times&\{(\xa-\xi)(\yb-\ya)\tfi+(\xf-\xi)(\yi-\ya)\t\}^{-1}
\non \\
&\times &\{(\xb-\xa)(\ya-\yb)\tfi+[2\xa(\yb-\yi)
\non \\
&+&2\xb(\ya-\yi)+(\xf+\xi)(2\yi-\ya-\yb)\t]\}^{-1}
\non \\
x_1^{~\prime}&=&\xb+(\xb-\xa)\biggl \{{(t-t_i)\over \t}
\non \\
&+&{[(\xa+\xb-2\xi)\tfi+(\xf-\xi)\t](\yb-\yi)\over
[(\xb-\xa)(\ya-\yb)\tfi+(\xf-\xi)(2\yi-\ya-\yb)\t]}\biggr \}
\non \\
y_1^{~\prime}&=&\yb+\biggr
\{(\xb-\xa)(\ya-\yb)(\yi-\yb)\tfi+\biggl [ [\xa(\yb-\yi)
\non \\
&+&\xb(\ya-\yi)](\yi+\yf-2\yb)+[\xf(\yi-\yb)+\xi(\yf-\yb)]
\non \\
&\times&(2\yi-\ya-\yb)
\biggr ]\t\biggr\}\biggl \{[(\xa+\xb-2\xi)\tfi+(\xf-\xi)\t](\yi-\yb)
\non \\
&+&[(\xb-\xa)(\yb-\ya)\tfi+(\xf-\xi)(\ya+\yb-2\yi)\t]{(t-t_i)\over \t}
\biggr \}
\non \\
&\times&\{(\xi-\xb)(\yb-\ya)\tfi+(\xf-\xi)(\yi-\yb)\t\}^{-1}
\non \\
&\times &\{(\xb-\xa)(\ya-\yb)\tfi+[2\xa(\yb-\yi)
\non \\
&+&2\xb(\ya-\yi)+(\xf+\xi)(2\yi-\ya-\yb)\t]\}^{-1}
\non \\
x_2^{~\prime}&=&\xf+{(x_f-x_i)\over\tfi}(t-t_f)
\non \\
y_2^{~\prime}&=&\yf+\{[(\xb-\xa)(\ya-\yb)\tfi+(\xf-\xi)(2\yi-\ya-\yb)\t]
\non \\
&\times &(\yf-\yi)(t-t_f)\}\{(\xb-\xa)(\ya-\yb)\tfi+[2\xa(\yb-\yi)
\non \\
&+&2\xb(\ya-\yi)+(\xf+\xi)(2\yi-\ya-\yb)\t]\tfi\}^{-1}
\lb{mirtraj}
\end{eqnarray}

Imposing the conditions \rff{327} for $\to$ and $\tb$ given by 
Eqs.~\rff{353}--\rff{144}, finally gives the following further constraints on
the trajectories
\begin{eqnarray}
0&>&\{\t(\xf-\xi)(2\yi-\yb-\yf)+\tfi[\xa(\yb-\yf)
\non \\
&+&\xb(2\ya-\yb-\yf)+2\xi(\yf-\ya)]\}
\non \\
&\times& [(\xb-\xa)(\ya-\yb)\tfi+(\xf-\xi)(2\yi-\ya-\yb)\t]^{-1}
\non \\
0&<&\{\t(\xf-\xi)(\ya-\yi)+\tfi[\xa(\ya+\yi-2\yb)
\non \\
&+&\xb(\yi-\ya)+2\xi(\yb-\yi)]\}
\non \\
&\times& [(\xb-\xa)(\ya-\yb)\tfi+(\xf-\xi)(2\yi-\ya-\yb)\t]^{-1}
\lb{352}
\end{eqnarray}
and
\begin{eqnarray}
0&<&{[(\xa-\xi)(\yb-\ya)\tfi+(\xf-\xi)(\yi-\ya)\t]
\over [(\xb-\xa)(\ya-\yb)\tfi+(\xf-\xi)(2\yi-\ya-\yb)\t]}
\non \\
0&>&{[(\xb-\xi)(\yb-\ya)\tfi+(\xf-\xi)(\yb-\yi)\t]
\over [(\xb-\xa)(\ya-\yb)\tfi+(\xf-\xi)(2\yi-\ya-\yb)\t]}
\lb{354}
\end{eqnarray}
which can be satisfied for
\begin{eqnarray}
0&<&(\xb-\xa)(\ya-\yb)\tfi+(\xf-\xi)(2\yi-\ya-\yb)\t
\non \\
0&>&\tfi[\xa(\yb-\yf)+\xb(2\ya-\yb-\yf)+2\xi(\yf-\ya)]
\non \\
&+&\t(\xf-\xi)(2\yi-\yb-\yf)
\non \\
0&<&\tfi[\xa(\ya+\yi-2\yb)+\xb(\yi-\ya)+2\xi(\yb-\yi)]
\non \\
&+&\t(\xf-\xi)(\ya-\yi)
\non \\
0&<&(\xa-\xi)(\yb-\ya)\tfi+(\xf-\xi)(\yi-\ya)\t
\non \\
0&>&(\xb-\xi)(\yb-\ya)\tfi+(\xf-\xi)(\yb-\yi)\t
\lb{355}
\end{eqnarray}
or for the analogous conditions with all the inequality signs reversed.

$\bullet$ {\it `Collinear velocities'}

For the `collinear velocities' solution for the conser\-va\-tion
Eqs.~\rff{319}, we can re\-write Eqs.~\rff{320}, using Eqs.~\rff{314}, to
explicitly show the dependence on $\ro, \to$ and $\tb$ as
\begin{eqnarray}
0&=&{\roi\over \toi}+{\roa\over \tob}
\non \\
0&=&{\rob\over \tbto}+{\rof\over \tfo}
\non \\
0&=&{\roi^2\over \toi^2}-{\rof^2\over \tfo^2}
\lb{356}
\end{eqnarray}
The algebra involved in the solution of the stationarity problem
defined by Eqs.~\rff{356} greatly simplifies if we choose to work in
the reference frame where the collision event is taken as the origin
of the spatial coordinates, i.e. $\ro~\dot=~(0, 0)$,
and the $x$-axis is taken along the direction defined by the (collinear)
velocities $\vuno$ and $\vdue$ (in other words, along the direction
of the vector $\ra-\ri$).
In this frame, the set of conditions \rff{356} is replaced by the much
simpler, equivalent system of equations
\begin{eqnarray}
0&=&\ya~=~\yi~=~\xo~=~\yo
\non \\
0&=&{\yb\over \xb}-{\yf\over \xf}
\non \\
0&=&{\xa\over \tob}+{\xi\over \toi}
\non \\
0&=&{\xb\over \tbto}+{\xf\over \tfo}
\non \\
0&=&{\xi^2\over \toi^2}-{(\xf^2+\yf^2)\over \tfo^2}
\lb{360}
\end{eqnarray}
We can easily solve the third and fourth of Eqs.~\rff{360} in terms of
$\to$ and $\tb$ as
\begin{equation}
\to={(t_f\xb\xi-t_i\xa\xf+\t \xi\xf)\over (\xb\xi-\xa\xf)}
\lb{362}
\end{equation}
\begin{equation}
\tb={[(t_f\xb+\t\xf)(\xa+\xi)-t_i\xa(\xf+ \xb)]\over (\xb\xi-\xa\xf)}
\lb{363}
\end{equation}

Moreover, using Eqs.~\rff{362}--\rff{363} into Eqs.~\rff{314},
we can find the classical trajectories as
\begin{eqnarray}
\xuno(t)&=&-\xdue(t)~=~
{[(t_f-t)\xb\xi+(t-t_i)\xa\xf+\t\xi\xf]\over [\tfi\xb+\t\xf]}
\non \\
\yuno(t)&=&\ydue(t)~=~0
\non \\
\runop(t)&=&-
{[(t_f-t)\xb\xi+(t-t_i)\xa\xf+\t\xi\xf]\over [\tfi\xa+\t\xi]\xb}\rb
\non \\
\rduep(t)&=&
{[(t_f-t)\xb\xi+(t-t_i)\xa\xf+\t\xi\xf]\over [\tfi\xa+\t\xi]\xf}\rf
\lb{368}
\end{eqnarray}
Besides the `coordinate-frame' constraints on these trajectories, which
are given by the first two of Eqs.~\rff{360}, and the constraint given
by the last of Eqs. ~\rff{360}, which using Eq. \rff{362}
becomes
\begin{equation}
| [\tfi\xa+\t\xi]\xf |=\sqrt{\xf^2 +\yf^2}| \tfi\xb+\t\xf |
\label{new}
\end{equation}
we also have the extra constraints 
which are implied by conditions \rff{327}, i.e.
\begin{eqnarray}
0&<&{(\xa+\xi)[\tfi\xb+\t\xf]\over (\xb\xi-\xa\xf)}
\non \\
0&>&{(\xb+\xf)[\tfi\xa+\t\xi]\over (\xb\xi-\xa\xf)}
\non \\
0&>&{\xa[\tfi\xb+\t\xf]\over (\xb\xi-\xa\xf)}
\non \\
0&<&{\xb[\tfi\xa+\t\xi]\over (\xb\xi-\xa\xf)}
\lb{365}
\end{eqnarray}
which can be solved by
\begin{eqnarray}
0&<&\xb\xi-\xa\xf
\non \\
0&<&(\xa+\xi)[\tfi\xb+\t\xf]
\non \\
0&>&(\xb+\xf)[\tfi\xa+\t\xi]
\non \\
0&>&\xa[\tfi\xb+\t\xf]
\non \\
0&<&\xb[\tfi\xa+\t\xi]
\lb{366}
\end{eqnarray}
(or the same formulas with all the inequality signs reversed, as in the
previous cases).

\subsection{The nature of the stationary points}
\label{App:Nature}

In order to study the nature of the stationary points found
in the collision cases, we have to evaluate the eigenvalues of
the Hessian ($H$) of $\hat S_{cl, c}$, where $\hat S_{cl, c}$
is a function of the variables $\ro, \to$ and $\tb$.

Taking second order derivatives of the classical action \rff{315}
with respect to $\xo, \yo, \to$ and $\tb$ we find
\begin{eqnarray}
{\pa^2 \hat S_{cl, c}\over \pa\xo^2}\bigg\v_{stat}&=&
{\pa^2 \hat S_{cl, c}\over \pa\yo^2}\bigg\v_{stat}~=~m\biggl[{1\over \toi}
+{1\over \tbto}
\non \\
&+&{1\over \tob}+{1\over \tfo}\biggr ]_{stat}>0
\non \\
{\pa^2 \hat S_{cl, c}\over \pa\to^2}\bigg\v_{stat}&=&m\left[{\roi^2\over
\toi^3}+{\rob^2\over \tbto^3}+{\roa^2\over \tob^3}+{\rof^2\over \tfo^3}
\right ]_{stat}>0
\non \\
{\pa^2 \hat S_{cl, c}\over \pa\tb^2}\bigg\v_{stat}&=&
-{\pa^2 \hat S_{cl, c}\over \pa\tb\pa\to}\bigg\v_{stat}=m\left[
{\rob^2\over \tbto^3}+{\roa^2\over \tob^3}\right ]_{stat}>0
\non \\
{\pa^2 \hat S_{cl, c}\over \pa\xo\pa\to}\bigg\v_{stat}&=&m\left[-{\xoi\over
\toi^2}+{\xob\over \tbto^2}-{\xoa\over \tob^2}+{\xof\over \tfo^2}\right
]_{stat}
\non \\
{\pa^2 \hat S_{cl, c}\over \pa\yo\pa\to}\bigg\v_{stat}&=&m\left[-{\yoi\over
\toi^2}+{\yob\over \tbto^2}-{\yoa\over \tob^2}+{\yof\over \tfo^2}\right
]_{stat}
\non \\
{\pa^2 \hat S_{cl, c}\over \pa\xo\pa\tb}\bigg\v_{stat}&=&m\left[
-{\xob\over \tbto^2}+{\xoa\over \tob^2}\right ]_{stat}
\non \\
{\pa^2 \hat S_{cl, c}\over \pa\yo\pa\tb}\bigg\v_{stat}&=&m\left[
-{\yob\over \tbto^2}+{\yoa\over \tob^2}\right ]_{stat}
\non \\
{\pa^2 \hat S_{cl, c}\over \pa\xo\pa\yo}\bigg\v_{stat}&=&0
\lb{369}
\end{eqnarray}

The Hessian of $\hat S_{cl, c}$ is a symmetric $4\times 4$ matrix with
entries given by the expressions \rff{369}.
The eigenvalues $\l$ of $H$ can be found in the standard way by solving for
the determinant
\begin{equation}
Det \left ({H\over m}-\l I\right )=0
\lb{55b}
\end{equation}
which gives the following characteristic equation for $\l$
\begin{equation}
\l^4+a_3\l^3+a_2\l^2+a_1\l +a_0=0
\lb{293}
\end{equation}
It can be checked (after a somewhat long and uninteresting algebra which we
omit
here) that, {\it for all the `velocity rules' given by
Eqs.~\rff{19}-\rff{320}}, the coefficients in 
Eq.~\rff{293} are such that $a_3, a_1 <0$ and
$a_2, a_0 >0$, which is enough to show that all eigenvalues $\l$ must be
positive.
Using well known theorems from analysis, we conclude
that the stationary point for the action $\hat S_{cl, c}$ is a minimum along
each of the 4-d hypersurfaces defined by the collision data of cases
{\it a - c}.

\end{document}